# Single-Beam Velocimetry with Dual Frequency Comb Absorption Spectroscopy

DAVID YUN[1,*], SCOTT C. EGBERT[1], AUGUSTINE T. FRYMIRE[1], SEAN C. COBURN[1], JACOB J. FRANCE[2], KRISTIN M. RICE[3], JEFFREY M. DONBAR[3], GREGORY B. RIEKER[1,**]

[1]*Precision Laser Diagnostics Laboratory, University of Colorado Boulder, Boulder, CO 80309, USA*
[2]*Innovative Scientific Solutions Incorporated, Dayton, OH 45459, USA*
[3]*U.S. Air Force Research Laboratory, Wright-Patterson AFB, OH 45433, USA*
*[david.yun@colorado.edu](mailto:david.yun@colorado.edu)*
**[greg.rieker@colorado.edu](mailto:greg.rieker@colorado.edu)*

**Abstract:** Laser absorption Doppler velocimeters use a crossed-beam configuration to cancel error due to laser frequency drift and absorption model uncertainty. This configuration complicates the spatial interpretation of the measurement since the two beams sample different volumes of gas. Here, we achieve single-beam velocimetry with a portable dual comb spectrometer (DCS) with high frequency accuracy and stability enabled by GPS-referencing, and a new high-temperature water vapor absorption database. We measure the inlet flow in a supersonic ramjet engine and demonstrate single-beam measurements that are on average within 19 m/s of concurrent crossed-beam measurements. We estimate that the DCS and the new database contribute 1.6 and 13 m/s to this difference respectively.

## 1. Introduction

Velocimetry of gases is essential for many different flow applications such as meteorology [1–3], combustion [4–6], and aerodynamics [7–9]. For many applications, it is desirable to have a velocimetry sensor with a simple experimental setup and high enough spatial resolution to capture gradients in the flow. One powerful velocimetry method is laser absorption spectroscopy (LAS). As an optical method, it does not disturb the flow and additionally does not require significant optical access (in most cases) or tracer particles as with other optical methods such as particle image velocimetry or laser Doppler velocimetry. In LAS, a laser beam propagating through the sample is used to retrieve several path-averaged flow parameters such as pressure, temperature, species mole fraction, and velocity along the laser line-of-sight. Accommodating a single laser beam through the sample requires minimal hardware and optical access. Additionally, the single laser beam can provide spatial resolution corresponding to the length and diameter of the beam. However, velocimetry with LAS traditionally utilizes two crossed, counterpropagating beams [10–14] which increases both hardware and optical access requirements as well as the spatial footprint of the measurement. The increased spatial footprint reduces the spatial resolution of the measurement to the plane incorporating the crossed beams. In the current work, we demonstrate single-beam velocimetry. It is enabled by a stable, GPS-referenced dual frequency comb spectrometer (DCS) and a new $H_2O$ vapor spectroscopic absorption database that produces models with low absorption-transition positional error at high temperatures.

LAS relies on measurements of the absorption of laser light through a sample at frequencies resonant with the quantum transitions of the constituent molecules. If the gas has a bulk velocity ($U$) and the laser beam probes the gas at an angle ($\theta$) with respect to the flow direction (Fig. 1), then the position of the measured absorption feature(s) in static flow ($\nu_o$) will be shifted to a new position ($\nu$) by the Doppler effect according to Eq. 1 (where $c$ is the speed of light). The Doppler shift is then used to derive velocity as shown in Eq. 2.

$$\nu - \nu_0 = \frac{U\nu_0 \sin\theta}{c} \quad (1)$$

$$U = \frac{c(\nu - \nu_0)}{\nu_0 \sin\theta} \quad (2)$$

The static line position is retrieved from a reference spectroscopic database such as HITRAN2020 [15]. However, an error (or drift) in the spectrometer frequency, $\epsilon_s$, or an error in the database line position, $\epsilon_{db}$, will impart an error in the derived velocity ($U'$) as shown in Eq. 5 (where from Eq. 4 to Eq. 5 we replace $\nu_0 + \epsilon_{db}$ with $\nu_0$ in the denominator since $\nu_0 \gg \epsilon_{db}$). Thus velocity measurements using a single laser beam/path are very susceptible to spectrometer frequency error and spectroscopic database error.

$$\nu - \nu_0 + \epsilon_{db} + \epsilon_s = \frac{U'(\nu_0 + \epsilon_{db})\sin\theta}{c} \quad (3)$$

$$U' = \frac{c(\nu - \nu_0 + \epsilon_{db} + \epsilon_s)}{(\nu_0 + \epsilon_{db})\sin\theta} \quad (4)$$

$$U' = U + \frac{c(\epsilon_{db} + \epsilon_s)}{\nu_0 \sin\theta} \quad (5)$$

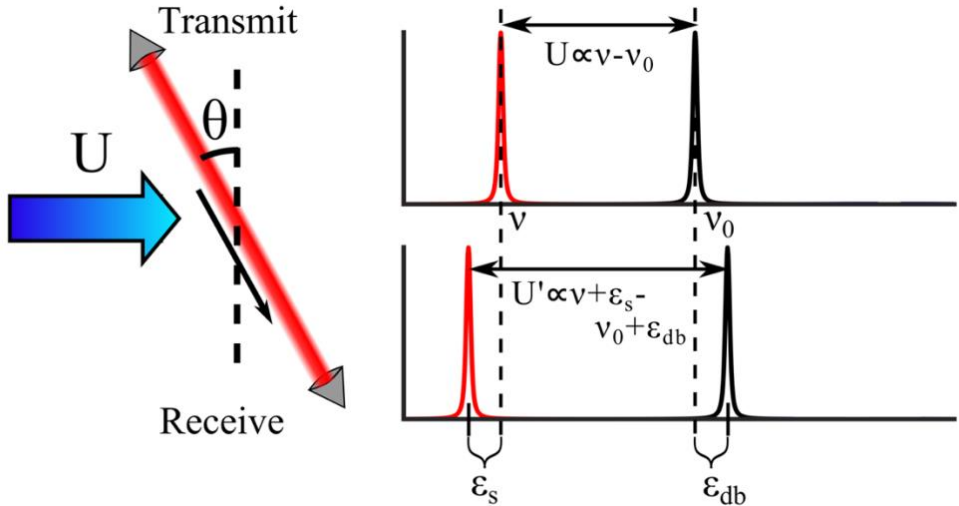

Figure 1. The left panel demonstrates a laser absorption spectroscopy setup for measuring velocity (U) with a single-beam configuration. A laser beam is sent at an angle (θ) to the normal of the bulk velocity which imparts a doppler shift to the absorption lines in the laser signal. An arrow indicates the direction of light propagation. The right plots demonstrate the Doppler shift for a single absorption line. In the ideal case (top plot) there is no error in the reference static line position ($\nu_0$) or in the measured line position ($\nu$) and the true velocity can be simply derived from the Doppler shift ($\nu - \nu_0$). However, in a realistic case (bottom plot) there can be an error in the database-derived reference static line position ($\epsilon_{db}$) and an error in the spectrometer-measured line position ($\epsilon_s$). Thus, the measured velocity (U') will be derived from an erroneous Doppler shift ($\nu + \epsilon_s - \nu_0 + \epsilon_{db}$) and will differ from the true velocity.

The sensitivity to spectrometer and database error led past researchers to utilize a crossed-beam configuration. If a second beam is counter-propagated at an equal but opposite angle to the first (see Fig. 2), the velocity can be determined from the difference in measured line positions between the spectra of the two beams ($\nu_1 - \nu_2$) as derived in Eq. 7 and shown in Fig. 2.

$$(v_1 - v_0) - (v_2 - v_0) = \frac{Uv_0 \sin\theta_1}{c} - \frac{Uv_0 \sin\theta_2}{c} \quad (6)$$

$$U = \frac{c(v_1 - v_2)}{v_0(\sin\theta_1 - \sin\theta_2)} \quad (7)$$

In this case, database and spectrometer positional errors effectively cancel when deriving velocity, as demonstrated below in Eq. 8-10.

$$(v_1 - v_0 + \epsilon_{db} + \epsilon_s) - (v_2 - v_0 + \epsilon_{db} + \epsilon_s) = \frac{U'(v_0 + \epsilon_{db})\sin\theta_1}{c} - \frac{U'(v_0 + \epsilon_{db})\sin\theta_2}{c} \quad (8)$$

$$(v_1 - v_2) = \frac{U'v_0(\sin\theta_1 - \sin\theta_2)}{c} \quad (9)$$

$$U' = U = \frac{c(v_1 - v_2)}{v_0(\sin\theta_1 - \sin\theta_2)} \quad (10)$$

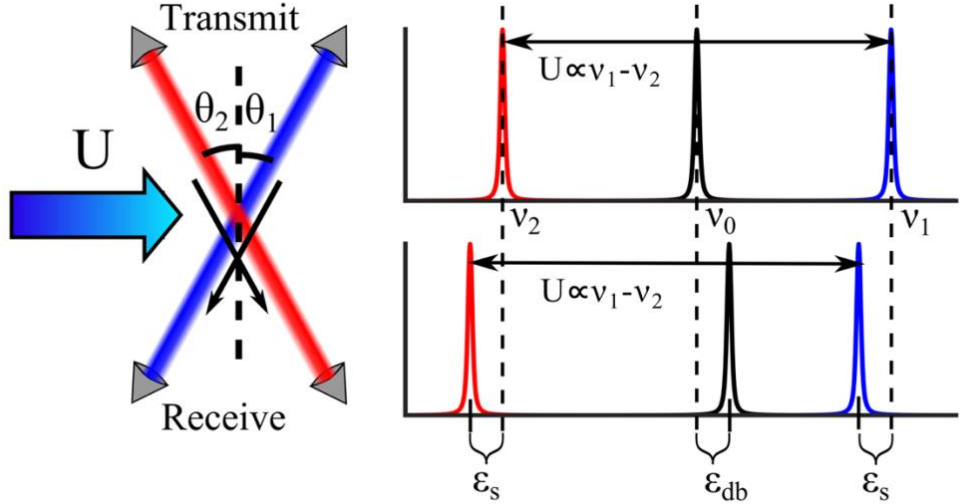

Figure 2. The left panel demonstrates a traditional laser absorption spectroscopy setup for measuring velocity (U) with a crossed crossed-beam configuration. Two laser beams are sent in counter-propagating directions (indicated by arrows) at angles $\theta_1$ and $\theta_2$ to the normal of the bulk velocity which create two spectra with opposite Doppler shifts to the absorption line in the laser signal. The right plots demonstrate the two Doppler shifts for a single absorption line. In the ideal case (top plot) there is no error in the reference static line position ($v_0$) or in the measured line positions ($v_1$ and $v_2$) and the velocity is derived from the difference in Doppler shifts ($v_1 - v_2$). In a realistic case (bottom plot) there can be an error in the database-derived reference static line position ($\epsilon_{db}$) and an error in the spectrometer-measured line position ($\epsilon_s$). However, we see that these errors don't affect the derived velocity as the difference between the two shifted line positions stays the same with the ideal case.

Thus, a crossed-beam measurement minimizes the effect of errors in both the spectrometer measurement and the absorption models (i.e., database error) by comparing the differential shift in laser absorption from the two beams [14]. However, a crossed-beam configuration requires additional experimental hardware (typically multiple sets of optics) and increases the spatial footprint of the measurement within the flow, thereby decreasing the spatial resolution and complicating the physical interpretation of the measurement with respect to CFD or other measurements. An alternative to the traditional crossed-beam setup is using two retroreflectors as demonstrated in Kurtz et al. [16]. Two retroreflectors facing each other create multiple passes across a sample region using a single laser. Each pass is at equal but opposite angles from the previous pass. This technique multiplies the pathlength and thus the SNR of the measurement while only requiring one set of transmit and receive optics. However, this setup also overlays the upstream and downstream Doppler-shifted absorption features on a single spectrum, which can impact the precision of the relative shift measurement if the features overlap (e.g. at lower velocity or higher pressure). The spatial footprint is also increased as the method results in an array of spatially dispersed beams.

Here, we demonstrate the first single-beam LAS velocimetry measurements. This is enabled through extensive work to minimize frequency errors stemming from both the spectrometer and the underlying spectroscopic database. In our implementation, we use a dual frequency comb spectrometer (DCS) that is clocked by a GPS-disciplined oscillator (GPSDO) and employs tight laser locking controls to realize low frequency drift ($\sim 2 \times 10^{-5}$ cm$^{-1}$). Additionally, we fit our data using a new near-IR $H_2O$ database that derives linecenters and pressure shifts from high-temperature $H_2O$-air laboratory measurements which we demonstrate to contribute a low frequency positional error in the model ($\sim 2 \times 10^{-4}$ cm$^{-1}$). We take our single-beam measurements of velocity in a ground-test ramjet isolator located at Wright-Patterson Air Force Base and validate against crossed-beam measurements taken concurrently with the same DCS.

## 2. Reducing measurement errors

A single-beam velocimetry measurement requires minimizing frequency error in the spectrometer, $\epsilon_s$, and frequency error in the absorption model line positions from the spectroscopic database, $\epsilon_{db}$. In this study, we reduce $\epsilon_s$ by using frequency comb lasers with highly accurate and stable frequency measurement and control. Frequency combs are specialized lasers that emit pulses of light at a fixed repetition rate that is often well known and controlled. In the frequency domain, these pulses comprise a broad spectrum of tightly spaced frequencies of light or "comb teeth" which have a frequency spacing equal to the pulse repetition rate. The position ($f_n$) for each comb tooth ($n$) is defined by the comb equation [17]:

$$f_n = nf_{rep} + f_0 \tag{11}$$

Thus, comb tooth positions are determined by the repetition rate ($f_{rep}$) and the comb offset ($f_0$) i.e., the carrier-envelope offset. To control the comb offset, we use the standard *f-2f* scheme where a frequency-doubled comb tooth from one end of the comb spectrum is heterodyned against a tooth from the opposite side of the spectrum to determine and stabilize the carrier-envelope offset [18–20]. To control the repetition rate of the laser, we lock one comb tooth to a continuous wave (CW) reference laser [21]. At the high comb tooth numbers for our measurements ($n \approx 1,000,000$), errors in repetition rate have a large effect on comb tooth positions. Thus, to minimize $\epsilon_s$, we must measure and control the repetition rate as accurately as possible.

In our system, we measure the repetition frequency with a detector and a digital frequency counter referenced to the GPSDO. The GPSDO has a 25 ppt relative frequency accuracy. At the comb frequencies used in this work (~7000 cm$^{-1}$), this frequency measurement accuracy results in a negligible contribution to $\epsilon_s$ of $1.75 \times 10^{-7}$ cm$^{-1}$ (7000 cm$^{-1} \times 25 \times 10^{-12}$). While the measurement of the repetition rate is accurate, there can still be drift in the repetition rate if the CW reference laser frequency is not controlled to maintain a stable f$_{rep}$. We use a control loop to minimize this drift by changing the current of the CW reference laser to maintain f$_{rep}$ at a specific value. Our current control loop and electronics maintain the frequency of the CW reference laser within $2 \times 10^{-5}$ cm$^{-1}$. Relative to the CW reference laser operating wavelength (6410 cm$^{-1}$), this is a frequency accuracy of 3 ppb and results in a value for $\epsilon_s$ of $2.2 \times 10^{-5}$ cm$^{-1}$ at the comb frequencies used in this work (~7000 cm$^{-1}$). Thus, the remaining CW laser drift is the main source of spectrometer frequency error.

A value of $\epsilon_s$ of $2.2 \times 10^{-5}$ cm$^{-1}$ results in a velocimetry error of 1.6 m/s (Fig. 3). This $\epsilon_s$ is a significant improvement over our previous DCS velocimetry works (Fig. 3). In previous works, we utilized a crossed-beam configuration to measure velocities with angles of 35° [14] and 12.5° [22] in part to overcome the absolute spectrometer drift. In our first work (Yun et al. 2022a [14]), we employed an oscillator that was not GPS-disciplined resulting in a value for $\epsilon_s$ of $1.1 \times 10^{-2}$ cm$^{-1}$ which leads to a velocity error of 750 m/s if a single-beam setup was used. For the second work (Yun et al. 2022b [22]), a GPSDO was employed but the drift of the CW reference laser was not as tightly controlled resulting in a $\epsilon_s$ of $1.8 \times 10^{-4}$ cm$^{-1}$ (velocity error = 35 m/s for a single beam configuration). The current system frequency drift contribution of 1.6 m/s velocity error to the single-beam measurement can be further improved with tighter locking of the CW laser frequency, but there are diminishing returns as other sources of velocity error become dominant.

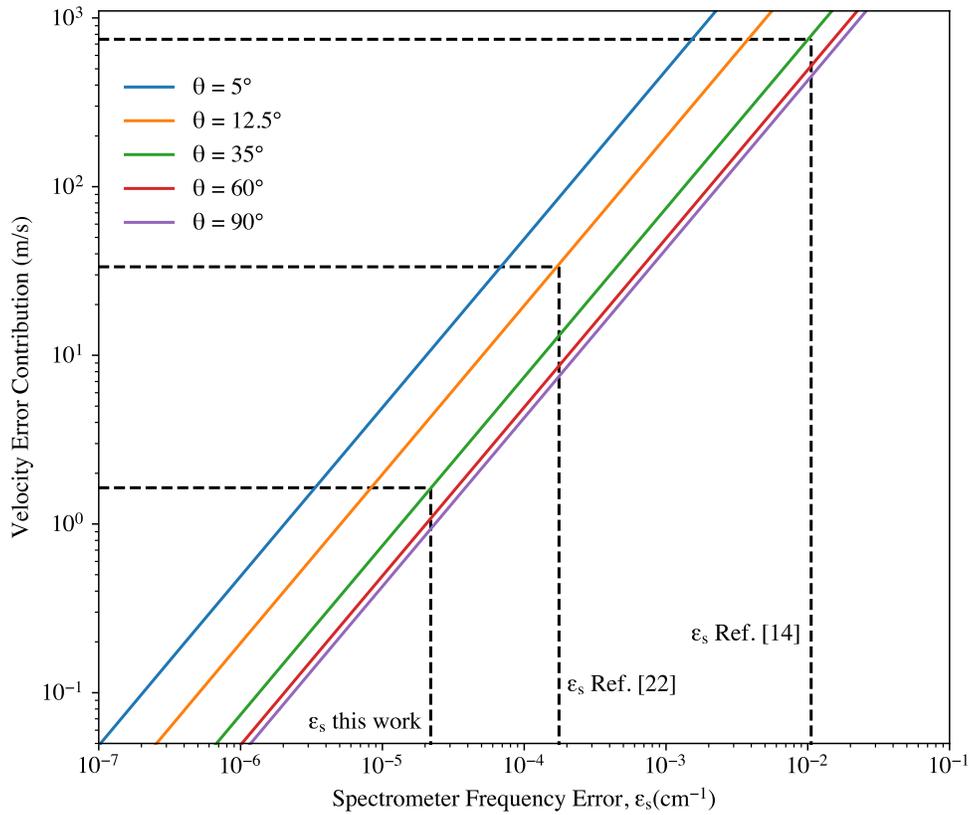

Figure 3. Plot demonstrating the relationship between frequency error in the spectrometer, $\epsilon_s$, and the resulting contribution to velocity error from a single-beam measurement at different beam angles to the flow normal (see Fig. 1). Dotted lines indicate the spectrometer frequency error for the DCS used in the current work and in our previous DCS velocimetry works and the resulting expected error contributions if these systems were used for single-beam velocimetry.

To minimize $\epsilon_{db}$, we employ a new near-IR $H_2O$ spectroscopic absorption database which updates database parameters that directly impact line positions, i.e., linecenters and pressure shift. The linecenter parameter corresponds to the absorption frequency of a transition in a zero-pressure and zero-velocity environment. The pressure shift parameter characterizes the change in linecenter from collision-induced pressure shift. Collisions between molecules change the spacing of quantum energy levels of the molecules leading to shifts in the line position. Pressure shift, $\delta$, is commonly modeled using either a linear relationship or a power law relationship. The commonly used spectroscopic database HITRAN2020 [15], employs the linear relationship as shown in Eq. 12.

$$\delta(P,T) = P \sum_x \chi_x [\delta_x + \delta_x'(T - 296)] \qquad (12)$$

Here, $\delta_x$ is the pressure shift coefficient for a collisional partner $x$ in the mixture and $\delta_x'$ is the temperature dependence of the shift. The total pressure shift is a summation of pressure shift contributions from each collisional partner, $x$, based on their respective molefractions, $\chi_x$. The linear relationship works best when multiple shift coefficients and temperature dependencies are derived for different temperature ranges [23]. However, currently HITRAN2020 only provides one pressure shift coefficient for each line optimized for 296 K and does not include temperature dependent shift parameters. Thus, we expect significant line position errors when applying HITRAN2020 to high temperatures. The new database used in this work based on Egbert et al. [24,25] utilizes the power law shown in Eq. 13.

$$\delta(P,T) = P \sum_x \chi_x \delta_x \left(\frac{296}{T}\right)^{n_x} \qquad (13)$$

The power law has been demonstrated to work well over large temperature ranges for infrared $H_2O$ transitions [26,27]. The new database derives pressure shift coefficients, $\delta_x$, and pressure shift temperature-dependent exponents, $n_x$, for the power law through a series of carefully controlled laboratory experiments measuring $H_2O$-air mixtures with pressures ranging from 0.0007 to 0.79 atm, temperatures of 300 to 1300 K, and water mole fractions of 0.02 to 1.

## 3. Experiment

We collected dual frequency comb spectroscopy measurements in a grounded, direct-connect dual-mode ramjet test facility at Wright Patterson Air Force Base. In our DCS, two frequency combs with slightly different repetition rates are mixed and the combined light is passed through the flow, after which it is collected onto a fast photodetector. The slight difference in repetition rates results in comb tooth position differences between the two combs that create a heterodyne signal on the photodetector in the radiofrequency (RF) domain [28]. This RF signal can be mapped back to the optical domain (THz region) based on the known optical frequency of each comb tooth (Eq. 11) and enables a direct conversion of the detected RF frequencies to an optical spectrum.

For this work, we use two erbium-doped fiber combs and combine their light using an optical fiber coupler. A single GPSDO and a single CW reference laser provide the frequency control for both combs (Fig. 4), which creates mutual phase coherence between them. After combining the light from the two combs, we spectrally window the DCS light to span 6600 – 7400 cm$^{-1}$ where $H_2O$ absorption is strong. We direct the light into the ramjet via quartz windows using a collimator angled at 36.6°. We send a second beam through the ramjet at an equal-but-opposite angle so we can derive crossed-beam values that serve as a reference for our single-beam values. The light traverses 8.5 cm across the diameter of the axisymmetric ramjet where it is coupled through a lens and collected into multimode fiber on the other side. The multimode fibers connect to fast photodetectors whose signals are digitized and recorded.

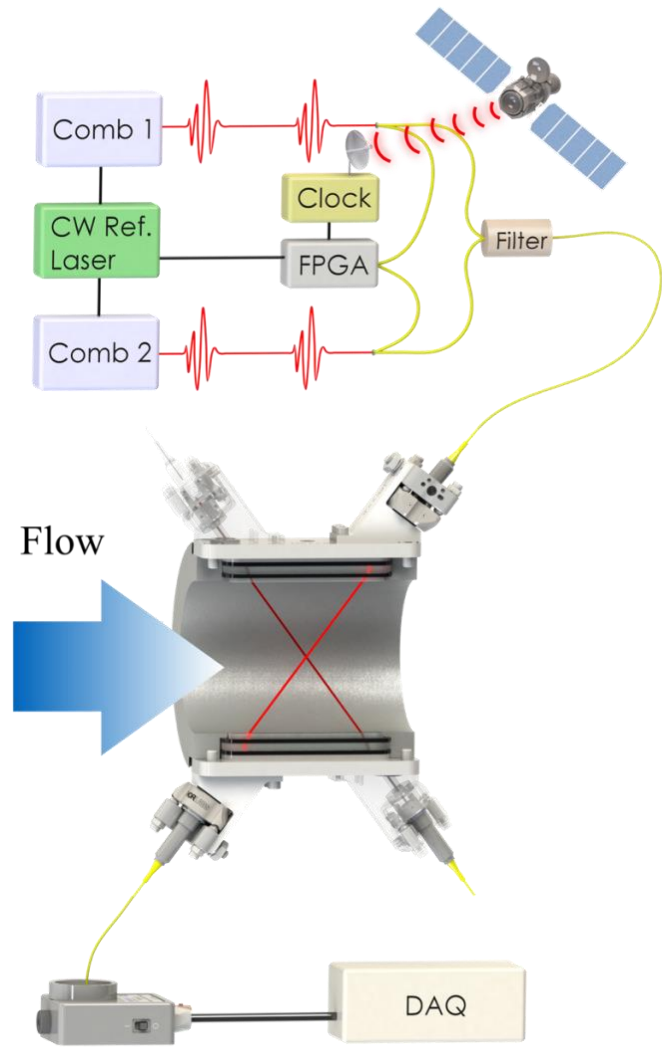

Figure 4. Schematic of DCS and experimental setup. The DCS setup consists of two frequency combs whose repetition rates are detected by an FPGA that is clocked by a GPS-disciplined oscillator. The FPGA makes corrections to the repetition rates via a control loop on a CW reference laser that is used to lock the comb teeth positions. Light from both combs is combined in a fiber coupler and then spectrally windowed with a fiber filter. The light is then sent through the ramjet flow at an angle to enable single-beam velocity measurements. Note - the light is also sent through a second set of optics (faded) at an equal but opposite angle to derive crossed-beam velocity values for validating the single-beam velocities. DCS light is measured with a photodetector and recorded on a data acquisition device (DAQ).

We measure five different flow conditions with velocities ranging from 900 – 1200 m/s. Spectra are acquired at 625 Hz and averaged for 150s for each condition. For this first demonstration of single-beam velocity measurements, we measure across simple flow conditions with large flat cores, small boundary layers, and radially symmetric flow to simplify the validation with the crossed-beam measurements. Throughout these experiments, we track both the timing error of our clock and the drift in the CW reference laser as shown in Fig. 5.

Timing error is collected from the GPSDO at approximately 30 second intervals and represents the time difference between a clock-generated 1 pulse-per-second source and the GPS signal (Fig. 5(b)). An Allan deviation of the timing error over 6000 s demonstrates that 150 s of averaging amounts to a 25 ppt frequency accuracy (Fig. 5(a)). The Allan deviation increases from 30s to 100s due to the slow ~100s drift in the timing error (Fig. 5(b)). Though we could get the same clock accuracy at 50 seconds averaging, we choose a longer averaging time to improve SNR in the laser absorption measurements to increase the velocimetry measurement precision.

Drift of the CW reference laser (Fig. 5(c)) is monitored by observing the difference between the repetition rate setpoint of the combs and the actual (measured) repetition rate for the duration of a measurement. The maximum drift of the CW reference laser across any of the measurements was $7 \times 10^{-5}$ cm$^{-1}$ (dashed lines) but the drift stayed within $2 \times 10^{-5}$ cm$^{-1}$ on average (solid lines). The average drift should correspond to the actual error on the reported velocity measurement, as the spectra themselves are averaged and then fit to determine the velocity-induced Doppler shift. Though we average our measurements for the full 150 s, the plot in Fig. 5(d) shows the instantaneous velocity error contribution from the DCS frequency error. The maximum instantaneous velocity error contribution from the spectrometer is 6 m/s (dashed lines) but stays within 1.5 m/s on average (solid lines).

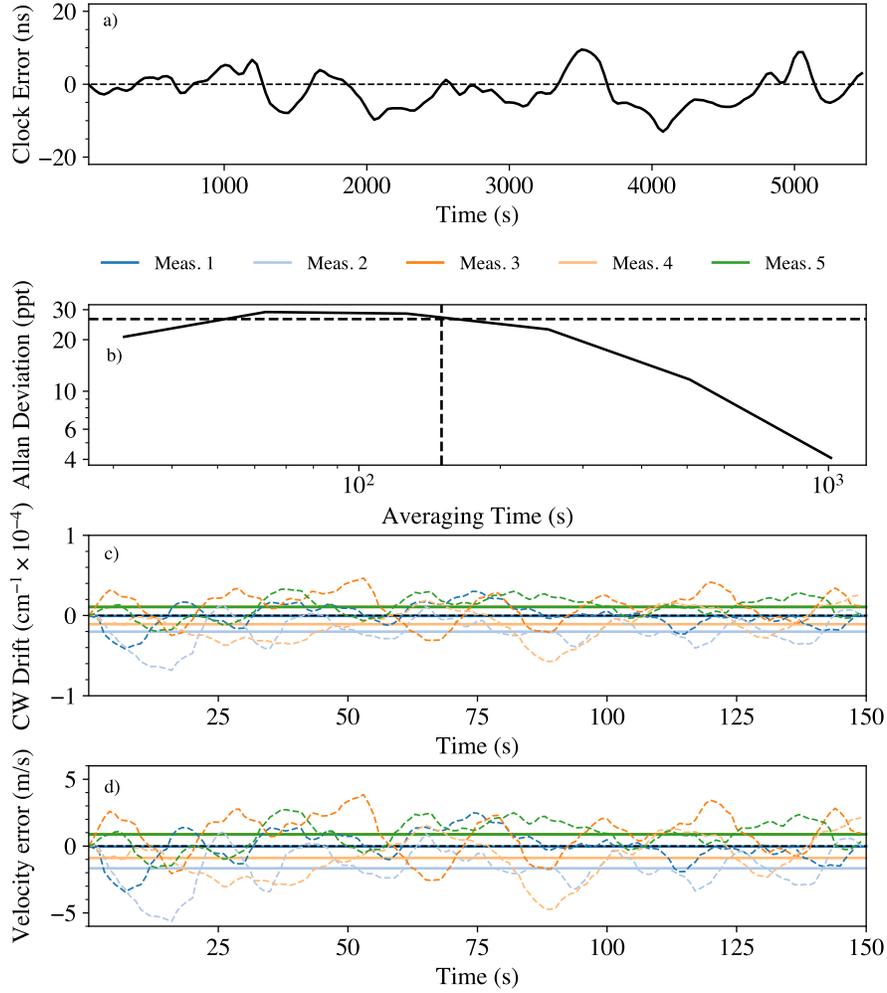

Figure 5. Performance of the DCS system during measurements. A plot of the clock timing error (a) shows that the timing error never exceeds 10 ns during the measurement times which are indicated by the color-shaded regions. An Allan deviation (b) demonstrates the clock frequency accuracy of 25 ppt at an averaging time of 150 s. A plot of the CW reference laser drift (c) during each measurement (color-coded by measurement number) demonstrates that the CW drift does not exceed $7\times10^{-5}$ cm$^{-1}$. When accounting for the clock accuracy and the CW drift we can predict the expected velocity error contribution of the DCS during each single-beam velocity measurement (d). Solid lines indicate the average values for a particular measurement, while dashed lines indicate instantaneous values in both panels (c) and (d).

## 4. Results

As with our past DCS velocimetry works [14,29], the recorded spectra are fit using modified free induction decay cepstral analysis [30]. Spectra are fit to absorption models derived from the Egbert *et al.* database [24,25] and we float pressure, $H_2O$ mole fraction, temperature, and velocity in the fits. For the single-beam retrievals, we fit spectra from the upstream-propagating beam and the downstream-propagating beam separately. An example fit from one of the downstream-propagating spectra is shown in Fig. 6(a-d). In order to validate the single-beam results, we also fit for crossed-beam velocities. In the crossed-beam fits, we input spectra from both beams to the fitting algorithm and constrain the fit to produce a single value for each parameter (pressure, $H_2O$ mole fraction, temperature, and velocity) across both spectra as done in Yun et al. [14]. A plot of the single-beam and crossed-beam velocities is shown in Fig. 6(e). For comparison, we also included a similar plot but with results generated using HITRAN2020 as the underlying spectroscopic database in Fig. 6(f). We chose HITRAN for this comparison because it is a commonly used database which has demonstrated more accurate spectra for near-IR $H_2O$ when compared to other databases [31].

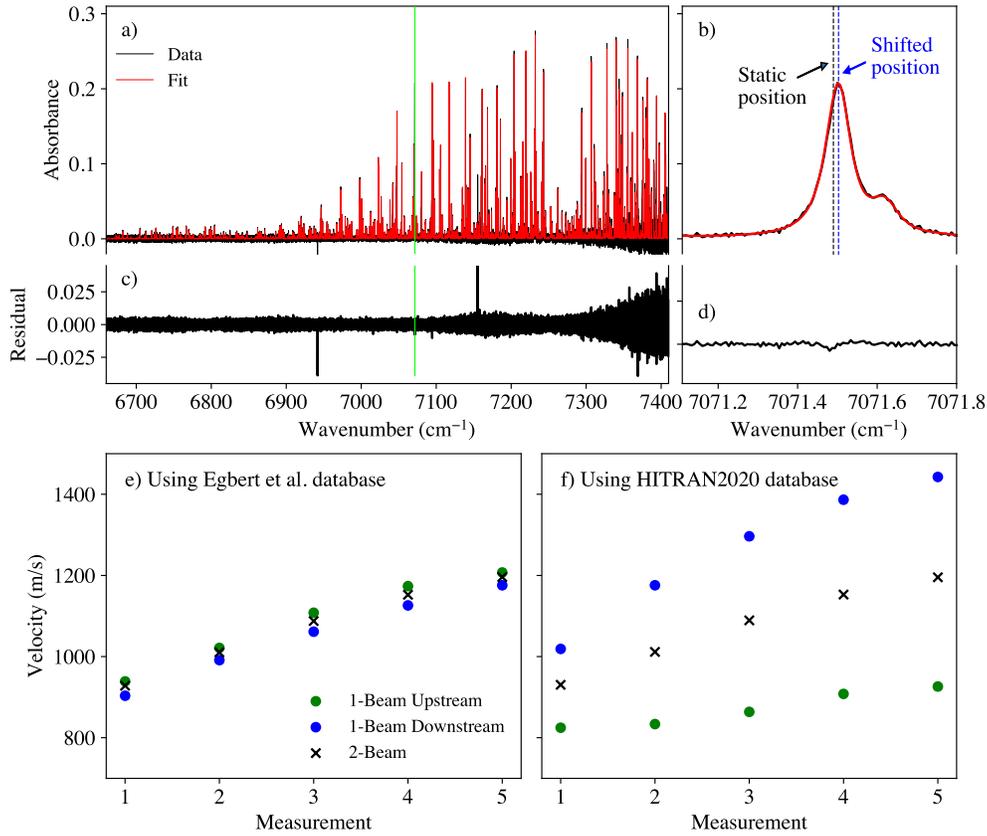

Figure 6. Results from a fit to example DCS absorbance data from the downstream-propagating beam of measurement #1 is shown in panel (a) where the DCS data is the black trace and the fit is the red trace. Panel b) shows a zoom view of this fit corresponding to the green region in panel (a) (7070.4 to 7072 cm$^{-1}$). The blue and black dashed lines show the doppler-shifted and static positions, respectively, of the line at 7071.5 cm$^{-1}$. Panel (c) shows the residual of the fit (data-fit) for the full fitting range (the two spikes are noise outliers) while panel (d) is the same but for the range from panel (b). The velocity results from the fit with the Egbert database for all five measurements are shown in panel (e) when fitting the upstream-propagating beam (green circle), the downstream-propagating beam (blue circle), and both beams (black x). Velocity results using HITRAN2020 as the database are similarly shown in panel f.

For results derived with the Egbert database, single-beam values differ from the crossed-beam values by 10 – 27 m/s (0.9 – 2.7% ) with an average of 19 m/s (1.8%). These correspond to shift errors ranging from $1.6 - 4.0 \times 10^{-4}$ cm$^{-1}$ with an average of $2.6 \times 10^{-4}$ cm$^{-1}$. Single-beam values derived with the HITRAN2020 spectroscopic database differ from crossed-beam values by 88 – 270 m/s (9 – 23%) with an average of 197 m/s (17.9%) corresponding to shift errors of $1.3 - 4.0 \times 10^{-3}$ cm$^{-1}$ (average = $2.7 \times 10^{-3}$ cm$^{-1}$). Most importantly, we can see that the velocity errors in the single-beam upstream and downstream measurements are of similar magnitude and in opposite directions. This suggests the majority of the error in the single-beam measurements derives from the database error (which would create this equal but opposite effect) and that the error from the Egbert *et al.* database is much lower compared to HITRAN 2020.

We perform an uncertainty analysis to quantify the contributions of the observed differences due to the experiment (i.e. the instrument, the optics setup) between the single-beam and crossed-beam velocity results to better understand the contribution of database error relative to other sources of uncertainty. Results from this analysis for measurement #2 are shown in Table 1.

Table 1. Experimental uncertainty for single-beam velocimetry for the upstream-propagating beam in measurement #2

|  | Shift (cm$^{-1}$) | Velocity (m/s) |
| --- | --- | --- |
| Spectrometer Frequency Error | $2.5 \times 10^{-5}$ | 1.8 |
| Angle Error | $4.8 \times 10^{-5}$ | 3.4 |
| Single-beam Precision | $4.9 \times 10^{-5}$ | 3.5 |
| Crossed-beam Precision | $3.5 \times 10^{-5}$ | 2.5 |
| Total Experimental Uncertainty | $7.0 \times 10^{-5}$ | 5.0 |

For this uncertainty analysis, we only consider sources that contribute to the difference between the single-beam and crossed-beam results; thus, we do not include some sources described in our previous DCS velocimetry works [14,29]. For instance, background absorption removal will affect both values equally and thus not contribute to a change in the difference between the single-beam and crossed-beam result, so is not included here. We consider here the spectrometer error, uncertainty in the beam angle, and precision due to noise in the spectral data. Spectrometer frequency error derives from frequency error in the clock and drift in the CW reference laser as discussed in the Sections 2 and 3 and imposes a 1.8 m/s or a corresponding frequency shift error of $2.5 \times 10^{-5}$ cm$^{-1}$ on average for measurement #2. In the experimental setup, we determine the beam angle to within 0.3° uncertainty for all runs, which can affect the velocity retrieval by 0.34% (Eq. 1) resulting in an uncertainty of 3.6 m/s ($5.4 \times 10^{-5}$ cm$^{-1}$). Precision due to measurement noise is determined by simulating spectra with noise characteristics similar to the data and deriving the scatter in the retrieved values from iteratively fitting these simulated spectra [32]. Precision is 3.5 m/s ($4.9 \times 10^{-5}$ cm$^{-1}$) for the single-beam velocity and 2.5 m/s ($3.5 \times 10^{-5}$ cm$^{-1}$) for the crossed-beam velocity. Precision is lower for the crossed-beam velocity because the crossed-beam measurement fit is based on twice as much spectral data [32]. To calculate the total uncertainty from the experiment, we sum the aforementioned uncertainties in quadrature.

We assume that the database error is the major unaccounted source after removing the total expected experiment uncertainty due to the above other sources from the difference between the single-beam and crossed-beam values and thus can estimate the database error as seen in Table 2. When fitting with Egbert, the remaining error ranges from 5.9 m/s ($8.8 \times 10^{-5}$ cm$^{-1}$) to 21.2 m/s ($3.2 \times 10^{-4}$ cm$^{-1}$) and is 14.1 m/s ($2.1 \times 10^{-4}$ cm$^{-1}$) on average. These errors are in line with the frequency uncertainty of the database itself (~$2 \times 10^{-4}$ cm$^{-1}$) [24]. For HITRAN2020, the remaining error ranges from 84.0 m/s ($1.3 \times 10^{-3}$ cm$^{-1}$) to 263.9 m/s ($4.0 \times 10^{-3}$ cm$^{-1}$) and is 191.4 m/s ($2.9 \times 10^{-3}$ cm$^{-1}$) on average. The larger errors from HITRAN2020 are expected since this database uses a linear pressure shift coefficient reference temperature of 296 K which works best over small temperature ranges, while the flows measured here range from 500 – 900 K.

Table 2. Calculation of contribution of database error to single-beam velocimetry for measurement #2

| | | Shift (cm$^{-1}$) | Velocity (m/s) |
|---|---|---|---|
| Egbert *et al.* | Difference between single-beam and crossed-beam result | $2.7 \times 10^{-4}$ | 19.3 |
| | Estimated Database Error | $2.0 \times 10^{-4}$ | 14.3 |
| HITRAN2020 | Difference between single-beam and crossed-beam result | $2.5 \times 10^{-3}$ | 177.9 |
| | Estimated Database Error | $2.4 \times 10^{-3}$ | 172.9 |

## 5. Conclusion

In this work, we demonstrate to our knowledge the first single-beam laser absorption velocimetry. We achieve this by using a mobile DCS system with accurate and stable GPS-based frequency referencing and by employing a new $H_2O$ spectroscopic absorption database that is derived from laboratory data taken over a large temperature range. The DCS system is updated from previous works with a GPS-disciplined oscillator and tighter control of the reference laser drift to reduce velocity error contribution by two orders of magnitude. The new $H_2O$ database derives pressure shift variables based on the power law to provide better model accuracy over large temperature ranges. Comparing fits with the new database to HITRAN2020, the new database produces single-beam values that improve agreement with the crossed-beam measurements by an order of magnitude. The new database produces single-beam values with an average difference of 19 m/s or $2.7 \times 10^{-4}$ cm$^{-1}$ with the crossed-beam values. By performing an uncertainty analysis, we determine that 13 m/s or $1.9 \times 10^{-4}$ cm$^{-1}$ of this difference is due to the database which is on par with the frequency uncertainty of the database itself; whereas the contribution from the spectrometer frequency referencing is only 1.6 m/s or $2.2 \times 10^{-5}$ cm$^{-1}$ on average.

### Funding

This research was sponsored by the Defense Advanced Research Projects Agency (W31P4Q-15-1-0011), Air Force Research Laboratory (FA8650-20-2-2418) and the Air Force Office of Scientific Research (FA9550-17-1-0224, FA8650-20-2-2418).

### Acknowledgements

We would like to thank the test operators at RC-18, Steve Enneking, Andrew Baron, and Justin Stewart, who made sure we had everything we needed to take these measurements. This work has been cleared by the Air Force under case number AFRL-2023-4805.